\documentclass{ws-procs975x65}            

\def\d{{\rm d}}

\def\s{{\rm s}}
\def\SM{{\rm SM}}
\def\dis{\displaystyle}
\def\nt{\notag}

\begin{document}
\title{
Deviation of yukawa coupling and Higgs decay in gauge-Higgs Unification
}

\author{Y. Adachi$^*$}

\address{Department of Sciences, Matsue College of Technology,\\
Matsue 690-8518, Japan\\
$^*$E-mail: y-adachi@matuse-ct.ac.jp
}

\author{N. Maru}

\address{ Department of Mathematics and Physics, Osaka City University,\\
				Osaka 558-8585, Japan\\	
E-mail: nmaru@sci.osaka-cu.ac.jp}

\begin{abstract}
We study the deviation of yukawa coupling in the gauge-Higgs unification scenario from the Standard Model one. 
Applying the obtained results to the tau and bottom yukawa couplings, 
 we numerically calculate the 
 signal strength of $gg\to H \to b\bar b , \tau \bar \tau$  
 in the gauge-Higgs unification. 
\end{abstract}


\bodymatter

\section{Introduction}
Gauge-Higgs unification (GHU) \cite{GH} is one of the attractive scenarios beyond the Standard Model(SM), 
which provides a possible solution to the hierarchy problem without supersymmetry \cite{HIL}. 
In this scenario, 
the SM Higgs boson and the gauge fields are unified into the higher dimensional gauge fields, 
{\em i.e.} Higgs boson is identified with extra spatial components of higher dimensional gauge fields.
Due to this, the quantum correction to Higgs mass is UV-finite
and calculable due to the higher dimensional gauge symmetry 
though the theory is the non-renormalizable
\cite{ABQ,MY}
.
The finiteness of other physical observables 
have been investigated 
\cite{LM,Maru,ALM}. 


The fact that the yukawa couplings are governed by the gauge principle
may deviate 
from the SM one\cite{GHYukawaflat,GHYukawaWarped}.
Actually, in the flat extra dimensional case,
the ratio of yukawa coupling of GHU and the SM one is derived as 
\begin{equation} 
				\label{f_GHU} 
\frac{f_{{\rm GHU}}}{f_{{\rm SM}}} \simeq \frac{g_4}{2}v \pi R \cot \left( \frac{g_4}{2} v\pi R \right)\, .
\end{equation} 
It is because the Higgs fields $A_y^{(0) } $
appears the Wilson line phase which is defined by 
 \begin{equation} 
W = P {\rm exp} \left[ i \frac{g}{2} \oint_{S^1} A_y \d y \right] 
= {\rm exp} \left[ ig_4 \pi R A_y^{(0)} \right],
\label{Wilson}
\end{equation} 
where $g, g_4$ are 5D and 4D gauge couplings, respectively. 
Namely,  $W$ is periodic 
with respect to $A_y$ under $A_y \to A_y+ 2/(g R)$ 
so that the yukawa couplings $f_{\rm GHU} $ in this scenario becomes periodic as shown in eq. (\ref{f_GHU}).  

Although the above result is general feature in such a scenario,
but their model\cite{GHYukawaflat}  is not realistic.
A crucial point is  that the brane mass terms of the fermions are not included. 
They are necessary for generating the flavor mixing\cite{flavorGHU} 
 and removing the exotic massless fermions absent in the SM. 
Therefore, it is important to study  the deviation of Yukawa coupling in a realistic model 
 to incorporate the appropriate brane mass terms. 

\section{Deviation of yukawa couplings in gauge-Higgs unification}
We consider an $SU(3) \times U(1)'$ GHU model in a five-dimensional flat space-time 
compactified on $S^1/Z_2$ with the radius $R$ of $S^1$. 
The up-type quarks except for the top quark, 
the down-type quarks and the charged leptons are embedded into ${\bf 3}$ and $\overline{{\bf 6}}$ 
representations of $SU(3)$, respectively \cite{SSS}. 
In order to realize the large top Yukawa coupling, 
the top quark is embedded into $\overline{{\bf 15}}$ representation of $SU(3)$ \cite{CCP}. 

The boundary conditions are assigned 
 to reproduce the SM fields as the zero modes. 
A periodic boundary condition with respect to $S^1$ 
 is taken for all of the bulk SM fields, 
 and the $Z_2$ parity is assigned for the gauge fields and fermions 
 in the representation ${\cal R}$ 
 by using the parity matrix $P={\rm diag}(-,-,+)$ in the following. 
 \begin{equation} 
A_\mu (-y) = P^\dag A_\mu(y) P, \quad A_y(-y) =- P^\dag A_y(y) P,  \quad 
\psi(-y) = {\cal R}(P) \gamma^5 \psi(y) 
\label{parity}
\end{equation}  
 where the subscripts $\mu$ ($y$) denotes the four (the fifth) dimensional component. 
With this choice of parities, 
zero-mode vector bosons in the model are only the SM gauge fields.


This parity assignment also leaves exotic massless fermions which is not included in the SM. 
Such exotic fermions are made massive 
 by introducing brane localized fermions with conjugate $SU(2) \times U(1)$ charges 
 and an opposite chirality to the exotic fermions, 
 allowing us to write brane-localized Dirac mass terms. 
These brane localized mass terms are also very important 
 to generate the flavor mixing in the context of GHU \cite{flavorGHU}. 
 

In this context, 
 the deviation of the tau and the bottom yukawa couplings in GHU from the SM one has been obtained as \cite{AM}
\begin{align}
		 \dis \frac{f_{\rm GHU}}{f_\SM}
	=&\dis
	\frac{M^2-m_{\tau(b)}^2}{M^2-\pi R m_{\tau(b)}^2\sqrt{M^2-m_{\tau(b)}^2}\coth(\pi R\sqrt{M^2-m_{\tau(b)}^2})}
	\nt	\\
	&
	\times
	\dis
	\pi RM_W
	\frac{\sin\left( 2\pi RM_W \right) - 
		\left[\sin\left( 2\pi RM_W \right)-\sqrt2 \sin\left(2\sqrt{2}\pi RM_W  \right)\right]\sin^2\theta}
		{1-\cos(2\pi RM_W) - \left[\cos(2 \sqrt 2 \pi RM_W)-\cos\left( 2\pi RM_W \right)\right]\sin^2\theta}.
\end{align}
The parameter $\theta$ is a mixing angle between two $SU(2)$ doublet zero modes existing per generation \cite{flavorGHU}. 
We can easily show that they reduce to the eq. (\ref{f_GHU}) 
or the result in the previous work\cite{GHYukawaflat} 
for the case  $\theta=0$.

\section{Calculation of the signal strength of $gg \to H \to b\bar{b}, \tau\bar{\tau}$}

In this section, we calculate the signal strength of $gg \to H \to b\bar{b}, \tau\bar{\tau}$. 
The Higgs production is dominated by the gluon fusion process at the LHC 
 and calculated from the coefficient of the following dimension five operator between the Higgs and the digluon, 
 \begin{equation} 
{\cal L}_{{\rm eff}} = C_{g} H G^a_{\mu\nu} G^{a\mu\nu} 
\end{equation} 
where $G_{\mu \nu}^a~(a=1-8)$ is the gluon field strength. 
 
%
In GHU,
we have to take into account the KK top loop contributions, 
 which is found to be
\begin{align}
	C_g^{\rm KK}=F(m_1) \times \frac{1}{2} \times 2 + F(m_2) \times 1 \times 2 + F(m_3) 
	\times \frac{3}{2} \times 1 + F(m_4) \times \frac{4}{2} \times 1
	\label{KKtop}
\end{align}
where the first factor behind $F(m_a)$ denotes the ratio for the top yukawa coupling and 
the second factor is a multiplicity of the same KK mass spectrum.    
\begin{align}
F(m_a) \equiv 
	-\frac{1}{16\pi}\frac{m_t}{v}\alpha_\s \sum_{n=1}^\infty
	\Bigg[&
		\frac{1}{m_{a+}^{(n)}}F_{1/2}\left( \left(\frac{2m_{a+}^{(n)}}{m_h} \right)^2 \right)
		- (+ \to -)
	\Bigg],
	\label{}
\end{align}
$m_t=2M_W$ and $(m_{a \pm}^{(n)})^2 = (n/R \pm a M_W)^2$.

We perform a numerical calculation of the deviation of the yukawa couplings 
and Higgs production $C^{{\rm KK}}_g/C^{{\rm SM}}_g$ 
.
Combining them, the signal strength $\mu$ of the process $gg \to H \to b\bar{b}, \tau \bar{\tau}$ 
\begin{equation} 
		 \mu = \left| \frac{C^{{\rm KK}}_g}{C^{{\rm SM}}_g} \right|^2 \times \left|  \frac{f_{\rm GHU} }{f_\SM} \right|^2 
\end{equation} 
are shown in Figure \ref{SS}. 
%
\def\scale{0.495}
\begin{figure}[h]
	\centering
	\includegraphics[scale=\scale]{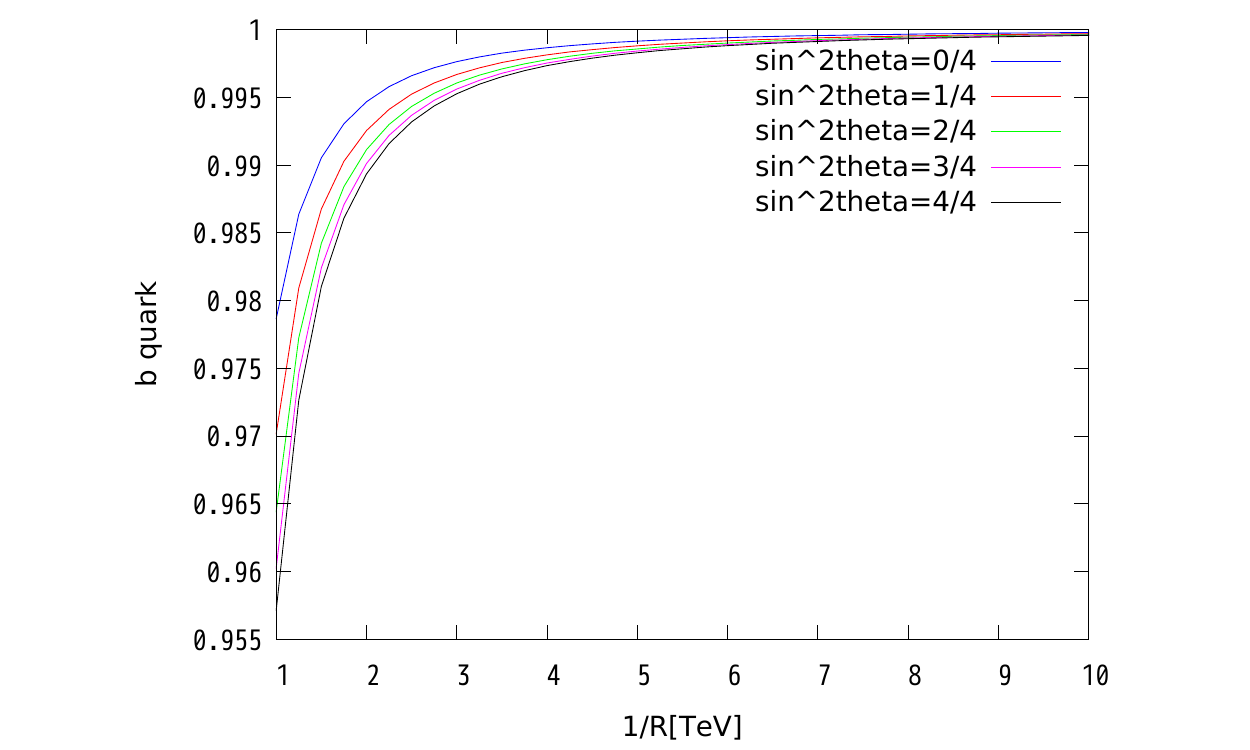}
	\includegraphics[scale=\scale]{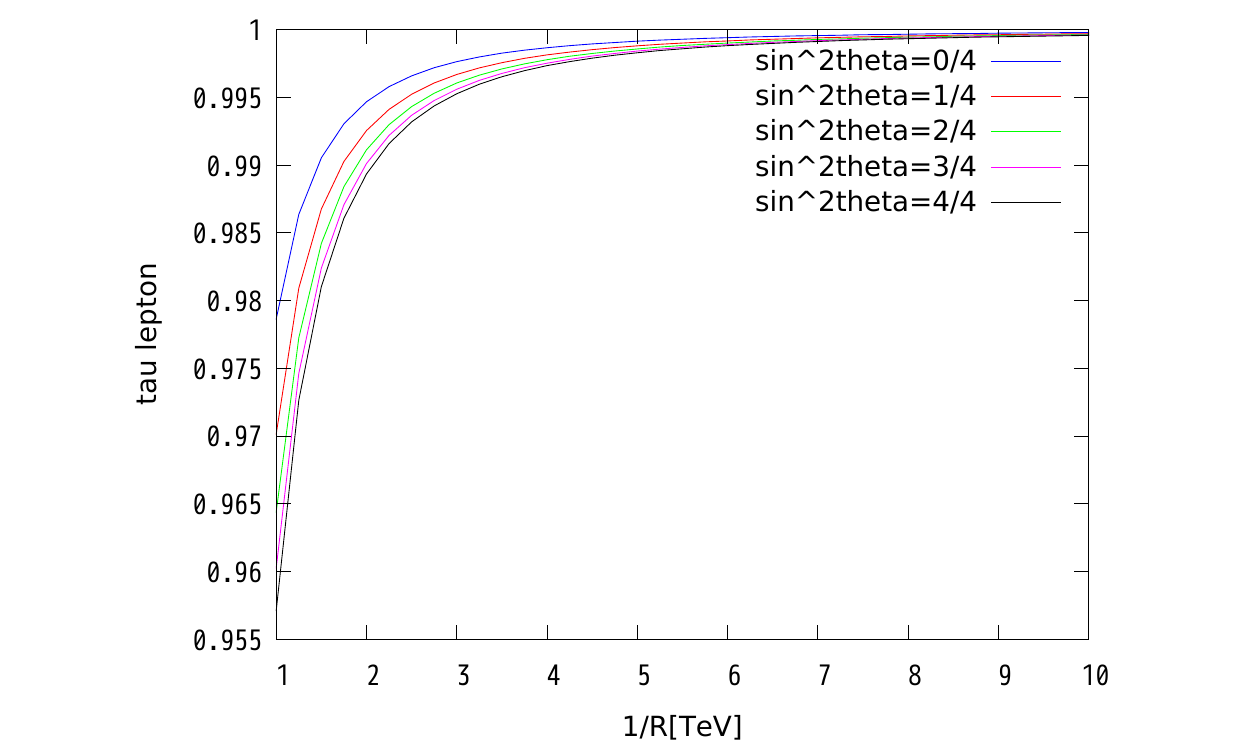}
	\caption{The left (right) plot is the signal strength of $gg \to H \to b\bar{b}(\tau \bar{\tau})$. 
	The horizontal line denotes the compactification scale. The results does not almost depend on the parameter $\theta$.}
	\label{SS}
\end{figure}
Our prediction is that the signal strength is always smaller than the unity, 
 namely the process $gg \to H \to b\bar{b}, \tau\bar{\tau}$ in GHU is always suppressed comparing to the SM prediction. 
 This is because the suppression is dominantly due to the suppression of the Higgs production via the gluon fusion\cite{MO}, 
 while the deviation of yukawa coupling is known to be very small \cite{AM}.   

\section{Summary}
In this presentation, 
we study the deviation of the yukawa coupling  in GHU scenario from SM one
and calculate 
the signal strength of the bottom and tau decays of the Higgs boson 
 produced via the gluon fusion at the LHC $gg \to H \to b\bar{b}, \tau\bar{\tau}$
 is suppressed.
Our generic prediction is that the deviation of yukawa coupling is quite small in the realistic parameter space 
and the signal strength is always smaller than the unity, 
 namely the process $gg \to H \to b\bar{b}, \tau\bar{\tau}$ in GHU is always suppressed comparing to the SM prediction. 
This is because the suppression is dominantly due to the suppression of the Higgs production via the gluon fusion, 
 while the deviation of yukawa coupling is known to be very small \cite{AM}.   
%


\begin{thebibliography}{10}


\bibitem{GH} 
  N.~S.~Manton,
  Nucl.\ Phys.\ B {\bf 158}, 141 (1979);
  D.~B.~Fairlie,
  Phys.\ Lett.\ B {\bf 82}, 97 (1979), 
  J.\ Phys.\ G {\bf 5}, L55 (1979);
  Y.~Hosotani,
  Phys.\ Lett.\ B {\bf 126}, 309 (1983), 
  Phys.\ Lett.\ B {\bf 129}, 193 (1983), 
  Annals Phys.\  {\bf 190}, 233 (1989).

\bibitem{HIL}
  H.~Hatanaka, T.~Inami and C.~S.~Lim,
  Mod.\ Phys.\ Lett.\ A {\bf 13}, 2601 (1998). 

\bibitem{ABQ}
  I.~Antoniadis, K.~Benakli and M.~Quiros,
  New J.\ Phys.\  {\bf 3}, 20 (2001); 
  G.~von Gersdorff, N.~Irges and M.~Quiros,
  Nucl.\ Phys.\ B {\bf 635}, 127 (2002); 
 R.~Contino, Y.~Nomura and A.~Pomarol,
  Nucl.\ Phys.\ B {\bf 671}, 148 (2003); 
  C.~S.~Lim, N.~Maru and K.~Hasegawa,
    J.\ Phys.\ Soc.\ Jap.\  {\bf 77}, 074101 (2008);  
  C.~S.~Lim, N.~Maru and T.~Miura,
  PTEP 2015 (2015) 4, 043B02.
  

\bibitem{MY}
  N.~Maru and T.~Yamashita,
  Nucl.\ Phys.\ B {\bf 754}, 127 (2006); 
  Y.~Hosotani, N.~Maru, K.~Takenaga and T.~Yamashita,
  Prog.\ Theor.\ Phys.\  {\bf 118}, 1053 (2007). 

\bibitem{LM}
  C.~S.~Lim and N.~Maru,
  Phys.\ Rev.\  D {\bf 75}, 115011 (2007). 

\bibitem{Maru}  
  N.~Maru,
  Mod.\ Phys.\ Lett.\  A {\bf 23}, 2737 (2008). 
  
\bibitem{ALM}
  Y.~Adachi, C.~S.~Lim and N.~Maru,
  Phys.\ Rev.\  D {\bf 76}, 075009 (2007); 
    Phys.\ Rev.\  D {\bf 79}, 075018 (2009);  
Phys.\ Rev.\  D {\bf 80}, 055025 (2009). 


 
\bibitem{GHYukawaflat} 
  K.~Hasegawa, N.~Kurahashi, C.~S.~Lim and K.~Tanabe,
  Phys.\ Rev.\ D {\bf 87}, 016011 (2013). 
  
\bibitem{GHYukawaWarped} 
  Y.~Hosotani and Y.~Kobayashi,
  Phys.\ Lett.\ B {\bf 674}, 192 (2009);
Y.~Hosotani, P.~Ko and M.~Tanaka,
  Phys.\ Lett.\ B {\bf 680}, 179 (2009). 


\bibitem{flavorGHU} 
  Y.~Adachi, N.~Kurahashi, C.~S.~Lim and N.~Maru,
  JHEP {\bf 1011}, 150 (2010); 
  JHEP {\bf 1201}, 047 (2012); 
   Y.~Adachi, N.~Kurahashi, N.~Maru and K.~Tanabe,
  Phys.\ Rev.\ D {\bf 85}, 096001 (2012); 
  Y.~Adachi, N.~Kurahashi and N.~Maru,
  arXiv:1404.4281 [hep-ph]. 





       
\bibitem{SSS} 
  C.~A.~Scrucca, M.~Serone and L.~Silvestrini,
  Nucl.\ Phys.\ B {\bf 669}, 128 (2003). 

\bibitem{CCP} 
  G.~Cacciapaglia, C.~Csaki and S.~C.~Park,
  JHEP {\bf 0603}, 099 (2006). 






\bibitem{AM}
Y.~Adachi and N.~Maru,
  arXiv:1501.4281 [hep-ph]. 
  






\bibitem{MO}
  N.~Maru and N.~Okada,
  Phys.\ Rev.\  D {\bf 77}, 055010 (2008);   
  Phys.\ Rev.\ D {\bf 87}, no. 9, 095019 (2013); 
  Phys.\ Rev.\ D {\bf 88}, no. 3, 037701 (2013); 
  arXiv:1310.3348 [hep-ph].   






\end{thebibliography}



\end{document}